\documentclass[sigconf]{acmart}

\usepackage{multirow}
\usepackage{listings}

\AtBeginDocument{%
  }

\copyrightyear{2026}
\acmYear{2026}
\setcopyright{cc}
\setcctype{by}
\acmConference[CHI EA '26]{Extended Abstracts of the 2026 CHI Conference on Human Factors in Computing Systems}{April 13--17, 2026}{Barcelona, Spain}
\acmBooktitle{Extended Abstracts of the 2026 CHI Conference on Human Factors in Computing Systems (CHI EA '26), April 13--17, 2026, Barcelona, Spain}
\acmDOI{10.1145/3772363.3798429}
\acmISBN{979-8-4007-2281-3/2026/04}

\begin{document}

\title{The Perfection Paradox: From Architect to Curator in AI-Assisted API Design}

\author{Mak Ahmad}

\orcid{0000-0001-8697-2035}
\affiliation{%
  \institution{Google}
  \streetaddress{}
  \city{San Francisco}
  \state{CA}
  \country{USA}
  \postcode{}
}
\email{seyeda@google.com}

\author{Andrew Macvean}
\orcid{0009-0003-8215-8513}
\affiliation{%
  \institution{Google}
  \city{Seattle}
  \country{USA}}
\email{amacvean@google.com}

\author{JJ Geewax}
\orcid{0009-0001-0153-5117}
\affiliation{%
  \institution{Google}
  \city{}
  \country{Singapore}}
\email{jjg@google.com}

\author{David Karger}
\orcid{0000-0002-0024-5847}
\affiliation{%
 \institution{MIT}
 \streetaddress{}
 \city{Cambridge}
 \state{MA}
 \country{USA}}
\email{karger@mit.edu}

\begin{abstract}
Enterprise API design is often bottlenecked by the tension between rapid feature delivery and the rigorous maintenance of usability standards. We present an industrial case study evaluating an AI-assisted design workflow trained on API Improvement Proposals (AIPs). Through a controlled study with 16 industry experts, we compared AI-generated API specifications against human-authored ones. While quantitative results indicated AI superiority in 10 of 11 usability dimensions and an 87\% reduction in authoring time, qualitative analysis revealed a paradox: experts frequently misidentified AI work as human (19\% accuracy) yet described the designs as unsettlingly ``perfect.'' We characterize this as a ``Perfection Paradox''---where hyper-consistency signals a lack of pragmatic human judgment. We discuss the implications of this perfection paradox, proposing a shift in the human designer's role from the ``drafter'' of specifications to the ``curator'' of AI-generated patterns.
\end{abstract}


\maketitle

\section{Introduction}

API design presents significant challenges at enterprise scale, where maintaining consistency across hundreds of services and distributed teams demands substantial coordination effort \cite{myers2016improving,robillard2009makes}. Studies show that API misuse remains widespread, with 88\% of applications containing at least one API usage mistake \cite{Eagle1}---errors often traceable to confusing API semantics and inconsistent design patterns rather than developer negligence \cite{Nadi1,Robilard1}. Poor API design leads not only to developer frustration but also to security vulnerabilities, integration failures, and increased maintenance costs that compound over time \cite{Xavier2017Historical,Bogart2016HowToBreak}.

Manual API governance---the process of reviewing APIs for style consistency and standard compliance---is inherently slow and prone to human error. While structured approaches like API Improvement Proposals (AIPs) and systematic design reviews have been shown to improve outcomes \cite{ahmad1,Macvean1}, these processes face fundamental scalability limitations. Research indicates that manual specification cycles require 2-4 hours per API \cite{murphy2}, and governance reviews struggle to keep pace with organizational growth as teams expand and API portfolios multiply \cite{ahmad1}. This bottleneck creates a difficult choice for organizations: either delay releases to ensure quality, or ship inconsistent APIs that accumulate technical debt. Large Language Models (LLMs) offer a compelling opportunity to address these challenges: they excel at pattern matching and can systematically apply established design rules with high consistency. Recent work has demonstrated substantial AI capabilities in code generation \cite{Kapitsaki,Yang2024AdvancingGA,Marques}, with AI matching human performance in well-defined programming tasks \cite{nascimento2023comparing}. This raises a fundamental question for software engineering practice: can we improve the API governance process by partnering humans with AI systems, allowing human designers to focus their expertise on domain complexity and strategic architectural decisions?

While prior work has explored AI-assisted API documentation \cite{dhyani2024automated}, specification inference \cite{decrop2024rest}, and testing \cite{pereira2024apitestgenie,sri2024automating}, none provide comprehensive design evaluation. Building on prior research examining AI-assisted API design workflows \cite{ahmad2}, we extend this investigation with focused expert evaluation, addressing three research questions:

\begin{itemize}
    \item \textbf{RQ1:} How do experts perceive AI-generated API designs compared to human alternatives across established usability dimensions?
    \item \textbf{RQ2:} Can experts distinguish AI-generated APIs from human-designed ones?
    \item \textbf{RQ3:} What API design tasks can AI effectively automate, and where does human expertise remain essential?
\end{itemize}

\section{API Generation System}
The AI system used GPT-4o, fine-tuned on Google's complete AIP specifications \cite{Google2024AIP}. These AIPs are formal documents articulating key API design considerations, serving as resources for human designers to ensure consistency. For this study, we adopt AIPs as our operational definition of ``good'' API design---a reasonable baseline given their widespread industry adoption and empirical validation in prior work \cite{ahmad1,ahmad2}. By fine-tuning on these specifications rather than using a general-purpose coding LLM, the goal was to create an ``API design expert'' that systematically encodes AIP-specific governance rules—patterns off-the-shelf models do not reliably apply—to generate consistent, high-quality definitions. Key implemented AIPs include resource modeling conventions (AIP-121, AIP-123), standard CRUD operations (AIP-131-136), pagination patterns (AIP-158), naming conventions (AIP-122, AIP-140), error handling (AIP-193), and long-running operations (AIP-151). Specialized prompts emphasized governance rule citation, consistent pattern application, and clear documentation generation \cite{arawjo2024chainforge,dunay2024multi}.

\begin{figure}[h]
  \centering
  \includegraphics[width=85mm,scale=1]{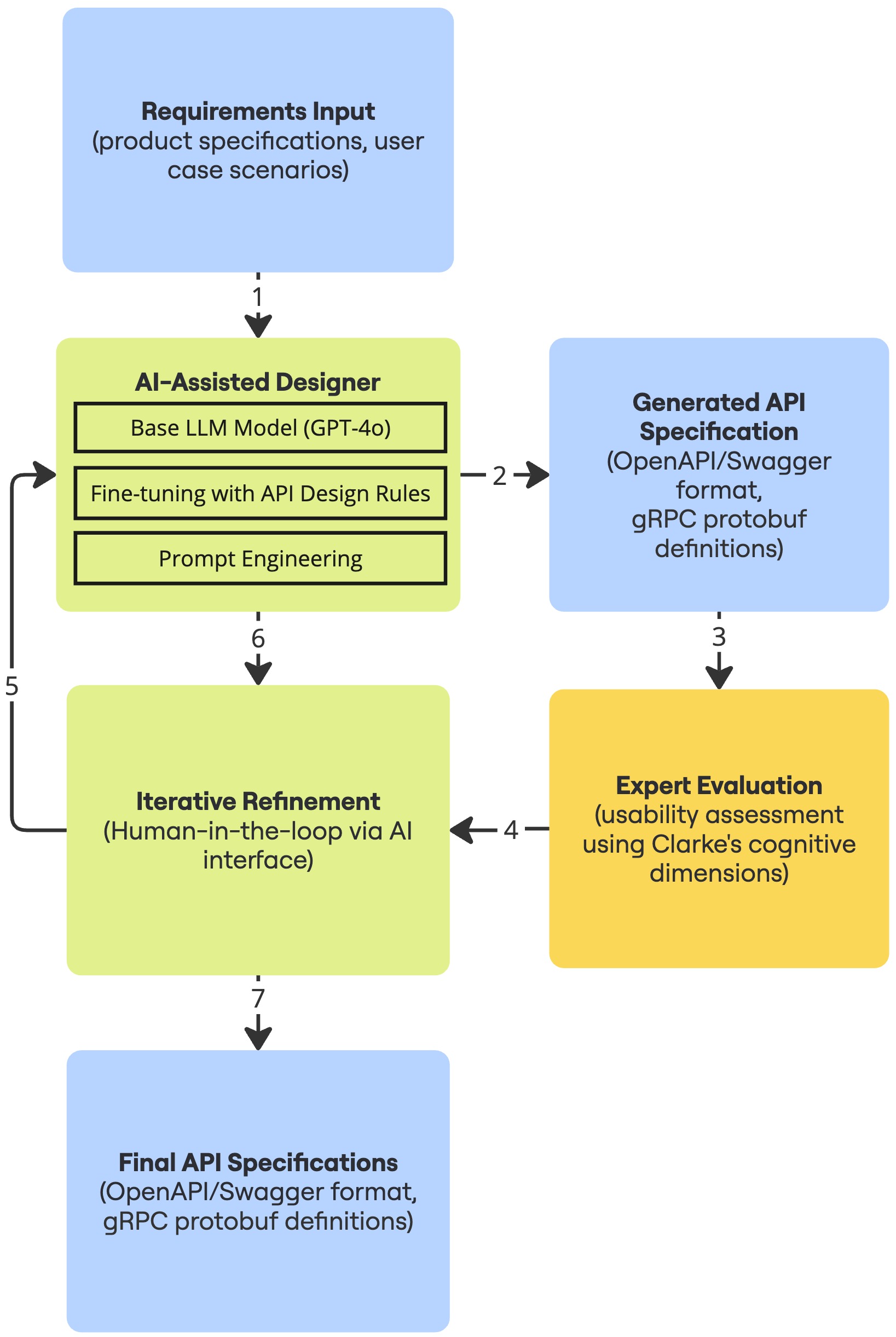}
  \caption{AI-Enhanced API Design Workflow: From Requirements to Final Specifications}
  \Description{This flowchart illustrates the AI-enhanced API design process. It consists of five main stages: Requirements Input, AI-Assisted Designer, Generated API Specification, Expert Evaluation, and Iterative Refinement. The process flows from top to bottom, with arrows indicating the sequence. A feedback loop connects Iterative Refinement back to the AI-Assisted Designer.}
  \label{fig:workflow}
\end{figure}

The generation process operates in three distinct phases (see Figure \ref{fig:workflow} for a detailed workflow diagram):
\begin{enumerate}
    \item \textbf{Requirement interpretation:} extracts functional requirements and maps user journeys to API operations.
    \item \textbf{AIP principle application:} ensures each endpoint follows governance standards.
    \item \textbf{Specification generation:} produces complete Protocol Buffer or OpenAPI definitions with inline documentation.
\end{enumerate}

Prior research on AI-human collaboration in software engineering \cite{Mili2024AIVH,martin2020ai,chong2024artificial} informed our design philosophy of complementing rather than replacing human expertise. The system is designed to handle routine governance enforcement while flagging complex design decisions for human review, creating a collaborative workflow that leverages the strengths of both AI consistency and human domain knowledge (see Figure \ref{fig:workflow} for a detailed workflow diagram).

\section{The Study}

\subsection{Task and Participants}

We selected a ``Children's Social Media Platform'' scenario because it requires meaningful architectural decisions beyond what automated tools like Swagger generators can infer from resource names and standard CRUD (Create, Read, Update, Delete) methods. APIs for such a system would require both standard CRUD operations and sensitive domain logic including parent-child relationships and content moderation. This task tests AI capabilities across complexity spectrums while representing enterprise patterns: hierarchical user management, content governance, and complex authorization. Senior architects from multiple industry sectors validated this task's industrial relevance through pilot testing before the main study commenced.

We recruited 16 API design experts through professional networks following established methodology \cite{Murphy}. Table~\ref{tab:demographics} summarizes participant characteristics. The study protocol was reviewed and approved by our organization's ethics board. To address documented concerns about evaluator background influence on AI artifact assessment \cite{MOSQUEIRAREY201846}, we collected detailed information about participants' prior AI tool experience. Following the practitioner perspective framework \cite{dos2024api}, we implemented multi-stage verification requiring documentation of past API design experience.

\begin{table}[t]
\caption{Participant Demographics (N=16)}
\label{tab:demographics}
\small
\begin{tabular}{p{0.36\columnwidth}p{0.56\columnwidth}}
\toprule
\textbf{Characteristic} & \textbf{Distribution} \\
\midrule
Programming Experience & 3--5y: 1, 6--10y: 2, 11--15y: 4, 16--20y: 2, 21+y: 7 \\
\midrule
API Design Experience & 1--2y: 1, 3--5y: 3, 6--10y: 4, 11--15y: 8 \\
\midrule
AIP Familiarity & Expert: 7, Very familiar: 2, Moderate: 1, Slight: 3, None: 3 \\
\midrule
AI Tool Experience & Extensive: 5, Moderate: 7, Limited: 4 \\
\bottomrule
\end{tabular}
\end{table}

\subsection{Evaluation Framework}

We employed Steven Clarke's Cognitive Dimensions of Notations framework \cite{Clarke1}, widely validated in HCI and software engineering for API usability assessment \cite{myers2016improving,murphy2,kelleher,Chamila}. This framework provides structured evaluation across 11 dimensions including Abstraction Level, Consistency, Penetrability, and Domain Correspondence, with an additional Requirements Fulfillment dimension. Prior studies \cite{stylos2,Marco1,stylos3,bernardo2020designing} demonstrate these dimensions effectively capture critical API design qualities.

\subsection{Study Design and Quality Control}
Participants evaluated three API specifications in blind fashion: one AI-generated and two human-designed specifications (drawn from two distinct design conditions in prior research \cite{ahmad2}, authored by experienced professional software engineers). Following the blind evaluation methodology established by Stylos and Clarke \cite{stylos2}, participants were not informed which specification was AI-generated until after completing all evaluations. Each specification was rated across all 11 dimensions using a 5-point Likert scale (1=Very Poor to 5=Excellent), with mandatory written justifications (minimum 25 words) per dimension. This requirement ensured participants engaged deeply with each specification rather than providing superficial assessments, and provided rich qualitative data for understanding rating patterns.

Quality control measures addressed concerns from prior expert evaluation studies \cite{Chamila,Rauf,Farooq1,grill,wijayarathna2018methodology}: response timing tracking flagged rushed completions, two attention validation questions per API verified engagement, and Latin square counterbalancing of presentation order controlled for order effects. The survey required approximately 30 minutes to complete. After completing all ratings, participants identified which specification was AI-generated and provided detailed reasoning (minimum 50 words). We conducted follow-up interviews with 8 participants selected through stratified sampling across experience levels and industry sectors. These 45-minute semi-structured discussions explored evaluation rationales and factors influencing identification decisions. Interview analysis used systematic thematic coding with two independent coders achieving 0.85 inter-rater reliability (Cohen's kappa).

\section{Results}

\subsection{AI Performance (RQ1)}

AI-generated APIs significantly outperformed human designs across 10 of 11 cognitive dimensions. Statistical analysis using Wilcoxon Signed-Rank tests confirmed these differences: Z = -3.517, $p < 0.001$ comparing AI (API B) versus Human A; Z = -3.062, $p = 0.002$ comparing AI versus Human C. Table~\ref{tab:results} presents mean scores across all dimensions. No statistically significant differences were found between the two human-authored specifications (Human A vs. Human C), confirming they serve as comparable baselines.

\begin{table}[h]
\caption{Mean Scores Across Cognitive Dimensions}
\label{tab:results}
\small
\begin{tabular}{lccc}
\toprule
\textbf{Dimension} & \textbf{Human (A)} & \textbf{AI (B)} & \textbf{Human (C)} \\
\midrule
Abstraction Level & 3.00 & 3.38 & 3.13 \\
Work-Step Unit & 3.06 & 3.19 & 3.19 \\
Working Framework & 3.19 & 3.50 & 3.13 \\
Progressive Evaluation & 3.25 & 3.31 & 3.00 \\
Premature Commitment & 2.88 & 3.25 & 2.63 \\
Penetrability & 3.06 & \textbf{3.75} & 2.69 \\
API Elaboration & 2.69 & 3.25 & 2.69 \\
Consistency & 3.06 & \textbf{3.50} & 3.13 \\
Role Expressiveness & 2.81 & 3.44 & 2.50 \\
Domain Correspondence & 2.81 & 3.38 & 2.88 \\
Requirements Fulfillment & 2.00 & \textbf{3.19} & 2.31 \\
\bottomrule
\end{tabular}
\end{table}

The strongest advantages appeared in three dimensions: Consistency (3.50 vs 3.06, Cohen's d=0.82), Penetrability (3.75 vs 3.06, d=0.89), and Requirements Fulfillment (3.19 vs 2.00, d=0.94)---likely reflecting AIP rule adherence more than deeper architectural soundness (see Section 5). Listings~\ref{lst:auth-ai} and~\ref{lst:auth-human} illustrate the architectural differences contributing to these consistency advantages.

\begin{lstlisting}[caption={AI-generated unified family account pattern, following AIP-122 resource naming and nested resource structure.},label={lst:auth-ai},float,language=C,basicstyle=\ttfamily\scriptsize,breaklines=true,frame=single]
service FamilyService {
  rpc CreateFamilyAccount(CreateFamilyRequest)
      returns (FamilyAccount);
  rpc AddChildProfile(AddChildRequest)
      returns (ChildProfile);
}
message CreateFamilyRequest {
  ParentProfile parent = 1;
  repeated ChildProfile children = 2;
  FamilySettings settings = 3;
}
message FamilyAccount {
  string name = 1;  // AIP-122 resource name
  ParentProfile parent = 2;
  repeated ChildProfile children = 3;
  google.protobuf.Timestamp create_time = 4;
}
\end{lstlisting}

\begin{lstlisting}[caption={Human-designed separate endpoint pattern with flatter structure.},label={lst:auth-human},float,language=C,basicstyle=\ttfamily\scriptsize,breaklines=true,frame=single]
service AuthenticationService {
  rpc CreateParentAccount(CreateParentRequest)
      returns (ParentAccount);
  rpc CreateChildAccount(CreateChildRequest)
      returns (ChildAccount);
  rpc VerifyEmail(VerifyEmailRequest)
      returns (VerificationResponse);
}
message CreateParentRequest {
  string name = 1;
  string email = 2;
  string password = 3;
}
message CreateChildRequest {
  string parent_id = 1;
  string child_name = 2;
  google.protobuf.Timestamp birth_date = 3;
}
\end{lstlisting}

The AI pattern treats family account creation as a single atomic operation with nested resources (following AIP-122), while the human pattern uses separate endpoints that require multiple round-trips. This structural difference directly explains the higher Consistency scores. 

Correlation analysis revealed meaningful relationships: Consistency and Penetrability showed strong correlation ($\rho = 0.743$, $p = 0.001$), and Domain Correspondence correlated with Requirements Fulfillment ($\rho = 0.629$, $p = 0.009$). Inter-rater reliability across all evaluators was strong: ICC(2,k) = 0.83 (95\% CI: 0.77-0.89).

In post-study debriefing sessions where participants learned which specification was AI-generated, P3 (senior architect, 15 years experience) noted the AI's ability to ``seamlessly incorporate AIP standards'' as particularly impressive.

\subsection{API Identification (RQ2)}

Expert ability to identify AI authorship was low: only 19\% of participants (3 of 16) correctly identified the AI-generated API. Chi-square analysis confirmed this was significantly below chance performance ($\chi^2(1) = 6.750$, $p = 0.009$). The majority of experts (81\%) incorrectly attributed human authorship to the AI specification, suggesting production-ready quality for established patterns.

Those who misidentified the AI as human-designed frequently cited its ``systematic consistency'' (P2, P8, P11) and ``comprehensive coverage of edge cases'' (P4, P14) as evidence of experienced human craftsmanship. P8 (platform engineer, 10 years) explained: ``The attention to detail felt very deliberate, like someone determined not to repeat mistakes.''

The minority who correctly identified the AI-generated specification noted subtle indicators, describing the AI designs as ``too elegant'' or ``algorithmically perfect'' compared to the pragmatic compromises typically found in human-authored specifications.

\subsection{Speed}

The AI system generated complete specifications in approximately 15 minutes compared to 2 hours typically required by human designers---representing an 87\% reduction in specification time \cite{Wadinambiarachchi,beer2024examination,IEEE2024SalarySurvey}. Participants emphasized that time savings should enable deeper human engagement with complex design decisions: ``The goal isn't to replace designers but to free them from mechanical consistency checking'' (P6, enterprise architect).

\section{Discussion}

\subsection{The Role of the Human: Intuition}

While the AI excelled at implementing established patterns consistently, expert analysis revealed important limitations in domain-specific nuance. Several participants identified architectural choices in the AI-generated specification that demonstrated strong consistency but raised practical concerns rooted in their operational experience.

P5 (fintech specialist, 11 years) ``Encapsulating all interactions as a single RPC seems very strange from an implementation perspective. These are all very different actions with different backend capabilities, consistency requirements, and scaling characteristics.'' P9 (platform architect, 14 years) elaborated on similar concerns: ``The API treats likes and comments as simple CRUD operations, but in high-scale systems, you need careful consideration of consistency models and race conditions. Likes are typically eventually consistent while comments often need stronger guarantees. A single unified model can't capture these operational realities.''

The AI also missed implicit domain requirements that experienced designers might intuit from context. For example, the AI-generated specification lacked endpoints for parents to block specific content categories or to restrict which accounts their children could follow---features implicit in the ``parental monitoring'' requirement but not explicitly enumerated. These required explicit specification that was not present in the original requirements document, aligning with recent research on AI limitations in novel scenarios requiring domain inference \cite{chen2024survey,ni2024l2ceval,tam2024framework}.

\subsection{The Role of the Human: Curator}

Our findings suggest AI excels as a governance enforcer---producing high-quality baseline specifications---while human designers add value as ``curators'' who refine AI-generated patterns based on operational constraints and domain nuance. The strong correlation between Consistency and Penetrability ($\rho = 0.743$) provides empirical support: AI-generated consistency enables better API comprehension, creating a foundation human curators can refine based on domain constraints and architectural styles \cite{murphy3,Robilard1,Fielding}.

In this curator model, the critical human decision becomes choosing strictly \textit{which} AIPs to optimize for. Context-dependent tradeoffs---between consistency and performance, or abstraction and granularity---require human judgment informed by organizational priorities. As P6 (enterprise architect) stated: ``The AI yields a solid foundation that would take hours to create manually. This allows human expertise to focus on the critical decision of when to deviate from established patterns to address specific operational realities.'' This aligns with recent findings on human-AI collaboration where the focus shifts from raw creation to review and design discourse \cite{alami2025human,GOPFERT2024100383}.

\subsection{Implications for Tool Design}

If the human role is evolving from drafter to curator, our current IDEs---optimized for character-by-character text entry---are ill-suited for the task. As noted in evaluations of code generation tools, developers often struggle to understand and debug AI-generated content \cite{vaithilingam2022expectation}. Consequently, we need new interfaces that adhere to generative AI design principles \cite{weisz2024design}, highlighting deviations from business logic rather than just syntax to help curators identify operational bottlenecks.

Our identification results highlight what we term a ``Perfection Paradox'' in code generation. While traditional software engineering seeks to eliminate variability, our experts found the AI's hyper-consistency unsettling. P9 described the designs as ``almost mechanical,'' while P5 noted they were ``elegant but perhaps too elegant,'' lacking the pragmatic compromises human engineers typically make for performance or backward compatibility.

This perfection paradoxically introduces new risks. In visual design, excessive polish can seem artificial; in API design, it creates what we term an ``illusion of correctness''---81\% of experts attributed the AI specification to human authorship, yet qualitative analysis revealed substantive architectural concerns (e.g., inappropriate abstraction choices, missing implicit requirements) masked by surface-level polish. Because the specifications perfectly adhered to syntax and AIP governance rules (the surface features'' of quality), experts found it cognitively difficult to critique the underlying structural logic. The generated APIs looked so professional that they bypassed the initial heuristic filters humans use to spot bad code. This suggests that AI assistance changes the nature of human review work: while reviewers have always needed to catch systemic flaws, AI-generated specifications eliminate the obvious syntactic and consistency errors that previously served as early warning signals. Reviewers must now engage more deeply with domain logic from the outset, as surface-level quality no longer correlates with structural soundness.

\section{Limitations}

The single evaluation scenario may not capture the full diversity of API design contexts. The 16-participant sample, while meeting established standards, limits generalizability. The focus on Google's AIPs represents one governance approach among many \cite{murphy3}; results may differ with other frameworks. Future work should examine long-term maintenance of AI-generated APIs \cite{tam2024framework} and extend evaluation across diverse governance frameworks.

\section{Conclusion}

Our evaluation demonstrates that AI-generated APIs trained on governance standards match or exceed human designs across usability dimensions while reducing specification time by 87\%. The 19\% identification rate, significantly below chance, suggests AI-generated specifications have reached production-plausible quality for established patterns. This technology lowers barriers to creating high-quality, governed APIs---particularly valuable for organizations struggling to maintain consistency across distributed teams. Future work should develop interfaces helping human curators spot subtle ``domain logic'' errors that AI systems make, now that ``syntax'' and consistency errors are largely eliminated. The shift from architect to curator represents not a diminishment of human expertise, but a strategic refocusing on architectural decisions, domain nuances, and operational constraints where human judgment provides maximum value.

\bibliographystyle{ACM-Reference-Format}
\bibliography{main}


\end{document}